\documentclass[sigconf]{acmart}

\usepackage{booktabs} 
\usepackage{graphicx, epstopdf}
\usepackage{subfigure}
\usepackage{url}
\usepackage{multirow}
\usepackage{algorithm}
\usepackage{algorithmic} 
\usepackage{balance}
\usepackage{arydshln}

\begin{document}
%
\title{Click-Through Rate Prediction with the User Memory Network}
\author{Wentao Ouyang, Xiuwu Zhang, Shukui Ren, Li Li, Zhaojie Liu, Yanlong Du}
\affiliation{%
  \institution{Intelligent Marketing Platform, Alibaba Group}
  \city{Beijing, China}
}
\email{{maiwei.oywt, xiuwu.zxw, shukui.rsk, ll98745, zhaojie.lzj, yanlong.dyl}@alibaba-inc.com}

\begin{abstract}
Click-through rate (CTR) prediction is a critical task in online advertising systems. Models like Deep Neural Networks (DNNs) are simple but stateless. They consider each target ad independently and cannot directly extract useful information contained in users' historical ad impressions and clicks. In contrast, models like Recurrent Neural Networks (RNNs) are stateful but complex. They model temporal dependency between users' sequential behaviors and can achieve improved prediction performance than DNNs. However, both the offline training and online prediction process of RNNs are much more complex and time-consuming. In this paper, we propose Memory Augmented DNN (MA-DNN) for practical CTR prediction services. In particular, we create two external memory vectors for each user, memorizing high-level abstractions of what a user possibly likes and dislikes. The proposed MA-DNN achieves a good compromise between DNN and RNN. It is as simple as DNN, but has certain ability to exploit useful information contained in users' historical behaviors as RNN. Both offline and online experiments demonstrate the effectiveness of MA-DNN for practical CTR prediction services. Actually, the memory component can be augmented to other models as well (e.g., the Wide\&Deep model).
\end{abstract}

\copyrightyear{2019}
\acmYear{2019}
\setcopyright{acmcopyright}
\acmConference[DLP-KDD'19]{1st International Workshop on Deep Learning Practice for High-Dimensional Sparse Data}{August 5, 2019}{Anchorage, AK, USA}
\acmBooktitle{1st International Workshop on Deep Learning Practice for High-Dimensional Sparse Data (DLP-KDD'19), August 5, 2019, Anchorage, AK, USA}
\acmPrice{15.00}
\acmDOI{10.1145/3326937.3341258}
\acmISBN{978-1-4503-6783-7/19/08}



\settopmatter{printacmref=true}
\fancyhead{}

\maketitle

\section{Introduction}
Click-through rate (CTR) prediction is to predict the probability that a user will click on an item. It plays an important role in online advertising systems. For example, the ad ranking strategy generally depends on CTR $\times$ bid, where bid is the benefit the system receives if an ad is clicked by a user. Moreover, according to the common cost-per-click charging model, advertisers are only charged once their ads are clicked by users. Therefore, in order to maximize the revenue and to maintain a desirable user experience, it is crucial to estimate the CTR of ads accurately.

CTR prediction has attracted lots of attention from both academia and industry \cite{he2014practical,cheng2016wide,he2017neural,wang2017deep,zhou2018deep}.
For example, Logistic Regression (LR) \cite{richardson2007predicting} models linear feature importance. Factorization Machine (FM) \cite{rendle2010factorization} further models pairwise feature interactions and shows improved performance. In recent years, Deep Neural Networks (DNNs) are exploited for CTR prediction in order to automatically learn feature representations and high-order feature interactions \cite{zhang2016deep,covington2016deep}. To take advantage of both shallow and deep models, hybrid models are also proposed. For example, Wide\&Deep \cite{cheng2016wide} combines LR and DNN to improve both the memorization and generalization abilities of the model.
DeepFM \cite{guo2017deepfm} combines FM and DNN to further improve the model ability of learning feature interactions.

One major limitation of the aforementioned models is that they mainly consider each target ad independently, but cannot directly extract useful information contained in users' historical ad impressions and clicks. To fill this gap, Zhang et al. \cite{zhang2014sequential} use Recurrent Neural Networks (RNNs) to model the dependency on users' sequential behaviors for CTR prediction. Tan et al. \cite{tan2016improved} propose improved RNNs for session-based recommendations. These RNN-based models do achieve improved performance. However, state-of-the-art RNN cell structures like Long Short-Term Memory (LSTM) \cite{hochreiter1997long} and Gated Recurrent Unit (GRU) \cite{chung2014empirical} are recursive and complex, which makes both the offline training and the online prediction process time-consuming. Moreover, the data preparation for RNNs is also complex. These issues make the application of RNNs to practical CTR prediction services rather complex (will be illustrated in \S\ref{sec_mot}).

In this paper, we propose Memory Augmented Deep Neural Network (MA-DNN) for CTR prediction in online advertising systems. In particular, we augment a DNN with an external user memory component, memorizing high-level abstractions of what a user possibly likes and dislikes. As a result, the proposed MA-DNN achieves a good compromise between DNN and RNN. It is as simple as DNN, but has certain ability to exploit useful information contained in users' historical behaviors as RNN. Actually, our proposal is flexible that the memory component can be augmented to other models such as Wide\&Deep \cite{cheng2016wide} and DeepFM \cite{guo2017deepfm} as well.
Both offline and online experiments demonstrate the effectiveness of the memory augmented neural networks for practical CTR prediction services.

\section{Model Design}

\subsection{Problem Statement}

\begin{figure*}[!t]
\centering
\includegraphics[width=0.88\textwidth]{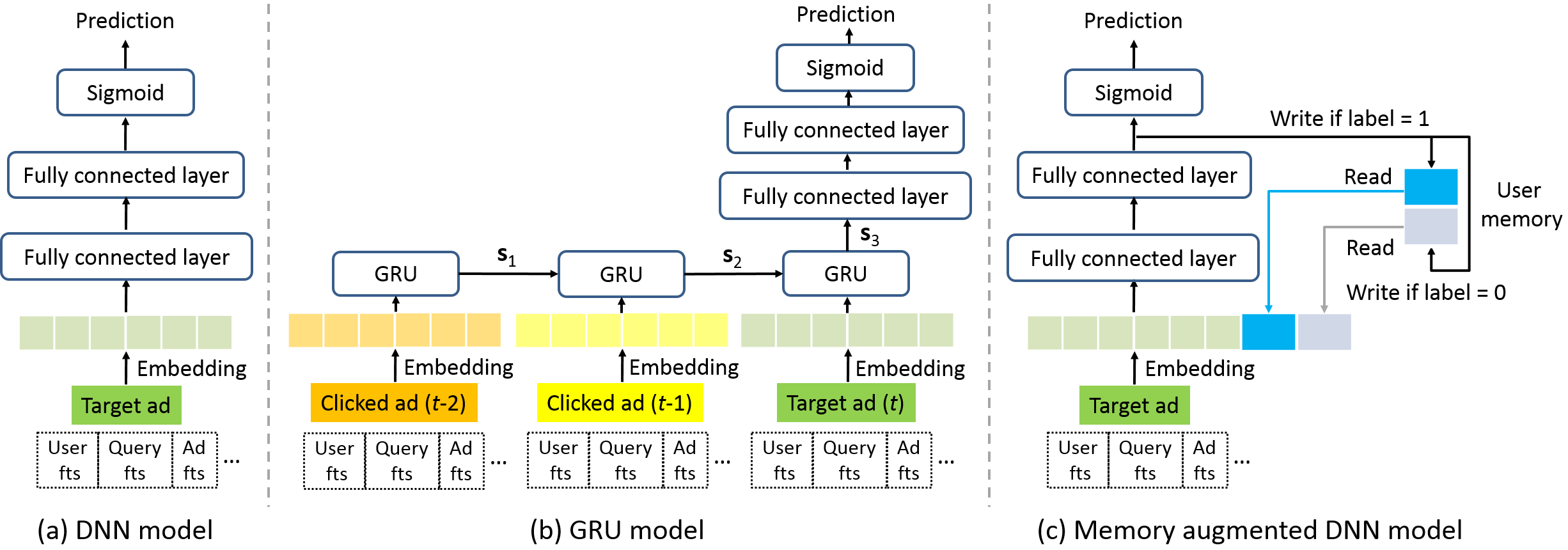}
\vskip -8pt
\caption{Structures of neural network models for CTR prediction.}
\vskip -8pt
\label{model}
\end{figure*}

\begin{table}[!t]
\small
\setlength{\tabcolsep}{3pt}
\renewcommand{\arraystretch}{1.2}
\caption{Each row is an instance for CTR prediction. The first column is the label (1 - clicked, 0 - unclicked). Each of the other columns is a field.}
\vskip -8pt
\label{tab_ft}
\centering
\begin{tabular}{|c|c|c|c|c|}
\hline
\textbf{Label} & \textbf{User ID} & \textbf{User Age} & \textbf{Ad Title} \\
\hline
1 & 2135147 & 24 & Beijing flower delivery \\
\hline
0 & 3467291 & 31 & Nike shoes, sporting shoes \\
\hline
0 & 1739086 & 45 & Female clothing and jeans \\
\hline
\end{tabular}
\vskip -12pt
\end{table}

The task of CTR prediction in online advertising is to estimate the probability of a user clicking on a specific ad.
Table \ref{tab_ft} shows some example instances. Each instance can be described by multiple \emph{fields} such as user information (User ID, City, etc.) and ad information (Creative ID, Title, etc.). The instantiation of a field is a \emph{feature}.

\subsection{Motivation}\label{sec_mot}
It is known that the DNN model is stateless. When DNN is used for CTR prediction, it cannot utilize useful information contained in past training instances. In contrast, the RNN model is stateful and it maintains a hidden state corresponding to each past training instance. It can thus achieve improved prediction performance. For example, Zhang et al. \cite{zhang2014sequential} use RNNs to model the dependency on users' sequential behaviors for CTR prediction.

Nevertheless, the application of RNNs to practical CTR prediction services is rather complex. We take GRU \cite{chung2014empirical}, one of the most advanced RNN cell structures, as an example. We analyze the complexity from two aspects: 1) model complexity and 2) data preparation complexity. We compare GRU with DNN.

\subsubsection{Model Complexity}
We illustrate the structures of DNN and GRU models for CTR prediction in Figure \ref{model}. The DNN model (Figure \ref{model}(a)) takes one ad instance as input.
The model contains an embedding layer, several fully connected layers and an output layer (i.e., the sigmoid function). Parameters in each fully connected layer only contain a weight matrix and a bias vector.

The GRU model (Figure \ref{model}(b)) takes a sequence of ad instances as input. Each instance first goes through an embedding layer and the resulting embedding vector $\mathbf{x}$ then goes through a GRU cell to result in a hidden vector $\mathbf{s}$. In particular, $\mathbf{s}_t = GRU(\mathbf{x}_t, \mathbf{s}_{t-1})$, i.e., the $t$th GRU cell takes the current instance embedding $\mathbf{x}_t$ and the hidden vector $\mathbf{s}_{t-1}$ from the last GRU cell as input. The final hidden representation ($\mathbf{s}_3$ in Figure \ref{model}(b)) goes through several fully connected layers and an output layer to result in the predicted CTR.

Each GRU cell is defined as
\begin{align}
& \mathbf{z} = sigmoid(\mathbf{x}_t \mathbf{U}^z + \mathbf{s}_{t-1} \mathbf{W}^z), \ \mathbf{r} = sigmoid(\mathbf{x}_t \mathbf{U}^r + \mathbf{s}_{t-1} \mathbf{W}^r), \nonumber \\
& \mathbf{h} = tanh(\mathbf{x}_t \mathbf{U}^h + (\mathbf{s}_{t-1} \circ \mathbf{r}) \mathbf{W}^h), \ \mathbf{s}_t = (1-\mathbf{z}) \circ \mathbf{h} + \mathbf{z} \circ \mathbf{s}_{t-1}, \nonumber
\end{align}
where $\circ$ denotes element-wise product, $\mathbf{x}_t$ and $\mathbf{s}_{t-1}$ are the input, $\mathbf{s}_t$ is the output and others are all model parameters.
We can clearly observe that GRU is much more complex than DNN. The complexity exists for both offline training and online prediction.

\begin{figure}[!t]
\centering
\includegraphics[width=0.36\textwidth]{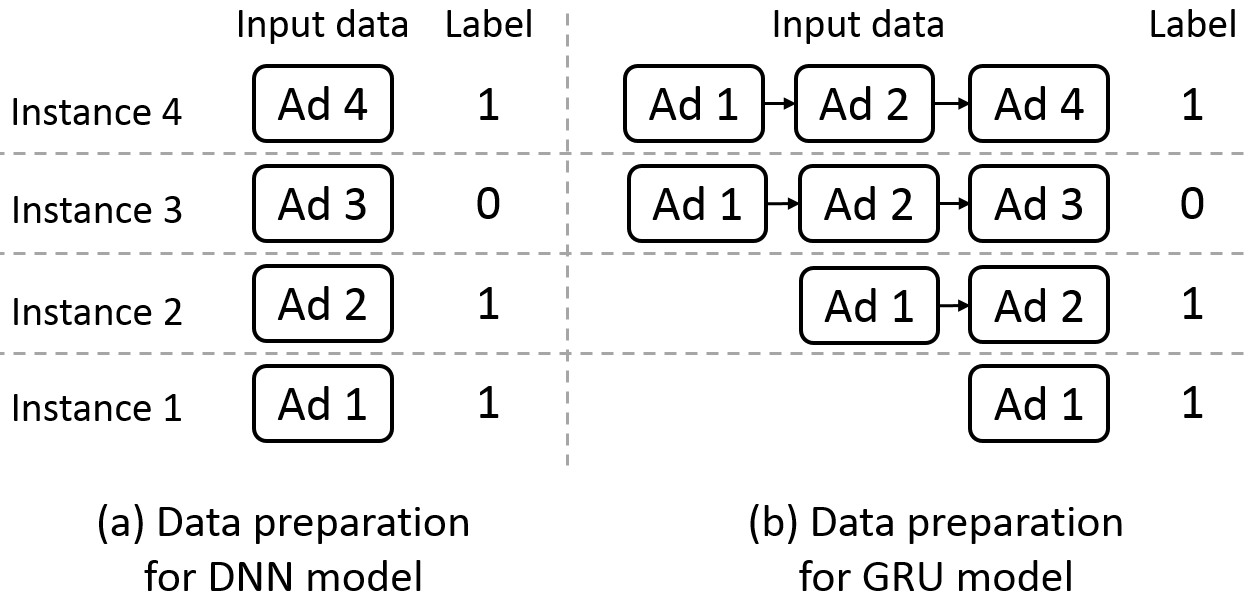}
\vskip -8pt
\caption{Illustration of data preparation.}
\vskip -12pt
\label{data_preparation}
\end{figure}

\subsubsection{Data Preparation Complexity}
We illustrate the data preparation complexity in Figure \ref{data_preparation}. It is observed in Figure \ref{data_preparation}(a) that DNN takes one ad instance as input. That is to say, it can directly use the log data where each ad impression is logged as a row.
In Figure \ref{data_preparation}(b), GRU takes a sequence of ad instances as input. Therefore, we have to search and concatenate the log data of a given user to form sequences as GRU input. Moreover, it is observed that the input data for GRU have clear redundancy. For example, ``Ad 1'' is repeated 4 times in 4 training instances. Nevertheless, data redundancy is not a problem for Natural Language Processing (NLP) tasks with GRU (e.g., sentence classification). This is because words in a sentence naturally form a sequence and there is no need to repeat words.


\subsection{Memory Augmented Deep Neural Network}
Based on the above analysis, we wonder whether we can design a new model that is as simple as DNN, but has certain ability to exploit useful information contained in users' historical behaviors as RNN.
To this end, we propose to augment the DNN model with an external user memory component. This component stores high-level information about a user's preferences. We illustrate the memory augmented DNN (MA-DNN) model structure in Figure \ref{data_preparation}(c).

\subsubsection{Model.} For a user $u$, we create two memory vectors $\mathbf{m}_{u1}$ and $\mathbf{m}_{u0}$, which memorize high-level information about what the user possibly likes and dislikes respectively.

First, each feature $f_i \in \mathbb{R}$ (e.g., a user ID) in an instance (whose CTR is to be predicted) goes through an embedding layer and is mapped to its embedding vector $\mathbf{e}_i \in \mathbb{R}^K$, where $\mathbf{e}_i$ is to be learned and $K$ is the embedding dimension. We then concatenate the embedding vectors from all the features as a long input vector $\mathbf{x}$.

Given the user ID $u$, we then find from the external memory the corresponding $\mathbf{m}_{u1}$ and $\mathbf{m}_{u0}$, and concatenate them with $\mathbf{x}$ to form a vector $\mathbf{v} = [\mathbf{x}, \mathbf{m}_{u1}, \mathbf{m}_{u0}]$.
Compared with DNN which takes $\mathbf{x}$ as input, the vector $\mathbf{v}$ not only contains information about current instance (i.e., $\mathbf{x}$), but also contains information about historical instances through the two memory vectors $\mathbf{m}_{u1}$ and $\mathbf{m}_{u0}$.

The vector $\mathbf{v}$ then goes through several fully connected (FC) layers with the ReLU activation function ($\mathrm{ReLU}(x) = \max(0, x)$), in order to exploit high-order nonlinear feature interactions \cite{he2017neural}. Nair and Hinton \cite{nair2010rectified} show that ReLU has significant benefits over sigmoid and tanh activation functions in terms of the convergence rate and the quality of obtained results.

Formally, the FC layers are defined as follows:
\begin{align}
\mathbf{z}_1 &= \mathrm{ReLU}(\mathbf{W}_1 \mathbf{v} + \mathbf{b}_1), \
\mathbf{z}_2 = \mathrm{ReLU}(\mathbf{W}_2 \mathbf{z}_1 + \mathbf{b}_2),  \ \cdots \nonumber \\
\mathbf{z}_L &= \mathrm{ReLU}(\mathbf{W}_L \mathbf{z}_{L-1} + \mathbf{b}_L),  \nonumber
\end{align}
where $L$ denotes the number of hidden layers; $\mathbf{W}_l$ and $\mathbf{b}_l$ denote the weight matrix and bias vector (to be learned) for the $l$th layer.

Finally, the output $\mathbf{z}_L$ of the last FC layer goes through a sigmoid function to generate the predicted CTR as
\[
\hat{y} = \frac{1}{1+\exp[- (\mathbf{w}^T \mathbf{z}_L + b)]},
\]
where $\mathbf{w}$ and $b$ are model parameters to be learned.

\subsubsection{Training.} We have two aims in the MA-DNN model: 1) we would like to make the predicted CTR as close to the true label as possible and 2) we would like the memory component to summarize what a user possibly likes and dislikes.

We achieve Aim 1 by minimizing the average logistic loss as
\begin{equation} \label{loss_1}
\mathrm{loss}_1 = - \frac{1}{|\mathbb{Y}|}\sum_{y \in \mathbb{Y}} [y \log \hat{y} + (1 - y) \log (1 - \hat{y})],
\end{equation}
where $y \in\{0,1\}$ is the true label of the target instance corresponding to $\hat{y}$ and $\mathbb{Y}$ is the collection of labels.

We achieve Aim 2 by minimizing the mean square error of the memory component and the output $\mathbf{z}_L$ of the last FC layer as
\begin{equation} \label{loss_2}
\mathrm{loss}_2 = \frac{1}{|\mathbb{Y}|}\sum_{y \in \mathbb{Y}} [y \mathbf{m}_{u1} + (1-y) \mathbf{m}_{u0} - \mathbf{z}_L]^2,
\end{equation}
where the user $u$ is identified by the user ID in the original input.
As can be seen, if the true label $y=1$, $\mathbf{z}_L$ is memorized by $\mathbf{m}_{u1}$ but not $\mathbf{m}_{u0}$; if the true label $y=0$, $\mathbf{z}_L$ is memorized by $\mathbf{m}_{u0}$ but not $\mathbf{m}_{u1}$. That is to say, $\mathbf{m}_{u1}$ and $\mathbf{m}_{u0}$ do summarize high-level information about what a user possibly likes and dislikes respectively by design. We choose to memorize the output $\mathbf{z}_L$ of the last FC layer because it contains the highest-level abstraction.

Since Aim 2 is to update $\mathbf{m}_{u1}$ and $\mathbf{m}_{u0}$ according to $\mathbf{z}_L$, we should not change the value of $\mathbf{z}_L$. To do so, the gradient of $\mathrm{loss}_2$ is only computed with respect to $\mathbf{m}_{u1}$ and $\mathbf{m}_{u0}$, but not $\mathbf{z}_L$. In this way, the value of $\mathbf{z}_L$ is updated only by $\mathrm{loss}_1$.

The final loss function is given by
$
\mathrm{loss} = \mathrm{loss}_1 + \alpha \mathrm{loss}_2,
$
where $\alpha$ is a tunable parameter.

%
%
%
%
%
%
%
%

\section{Experiments}

\subsection{Datasets}
Table \ref{tab_stat} lists the statistics of two large-scale experimental datasets.

1) \textbf{Avito advertising dataset\footnote{https://www.kaggle.com/c/avito-context-ad-clicks/data}.}
This dataset contains a random sample of ad logs from avito.ru, the largest general classified website in Russia.
We use the ad logs on 2015-05-20 for testing and others for training. No sampling is performed.
The features used include 1) ad features such as ad ID, ad title and ad category, 2) user features such as user ID, IP ID, user agent and user device, 3) query features such as search query, search category and search parameters, and 4) context features such as day of week and hour of day. The number of test samples is over $2.3 \times 10^6$.

2) \textbf{Company advertising dataset.}
This dataset contains a random sample of ad impression and click logs from a commercial advertising system in Alibaba. We use ad logs of 30 consecutive days in 2019 for training and logs of the next day for testing. The number of test samples is over $1.4 \times 10^6$.

\begin{table}[!t]
\small
\renewcommand{\arraystretch}{1.2}
\caption{Statistics of experimental large-scale datasets.}
\vskip -8pt
\label{tab_stat}
\centering
\begin{tabular}{|l|r|r|r|r|r|}
\hline
\textbf{Dataset} & \textbf{\# Training} & \textbf{\# Testing}  & \textbf{\# Fields} & \textbf{\# Features} \\
\hline
Avito & 168,255,929 & 2,332,738 & 27 & 42,301,586 \\
\hline
Company & 62,727,007 & 1,406,954 & 38 & 42,554,432 \\
\hline
\end{tabular}
\vskip -12pt
\end{table}

\subsection{Methods Compared}
\begin{enumerate}
\item \textbf{LR}. Logistic Regression \cite{richardson2007predicting}. It is a generalized linear model.
\item \textbf{FM}. Factorization Machine \cite{rendle2010factorization}. It models both first-order feature importance and second-order feature interactions.
\item \textbf{DNN}. Deep Neural Network. Its structure is in Figure \ref{model}(a). 
\item \textbf{W\&D}. The Wide\&Deep model in \cite{cheng2016wide}. It combines a wide component (LR) and a deep component (DNN). 
\item \textbf{MA-DNN}. Memory Augmented Deep Neural Network model. Its structure is shown in Figure \ref{model}(c).
\item \textbf{MA-W\&D}. Memory Augmented Wide\&Deep model.
\end{enumerate}

\subsection{Settings}
\textbf{Parameter Settings.}
The embedding dimension of each feature is set to $K=10$. The number of FC layers in neural network-based models is set to 3, with dimensions 512, 256 and 64. The dimensions of memory vectors $\mathbf{m}_{u1}$ and $\mathbf{m}_{u0}$ for a user $u$ are both set to 64.
The batch size is set to 128 and the balancing parameter $\alpha$ is set to 1. No dropout is performed.
Samples are fed to models in the time order  (i.e., no data shuffling).
All the methods are implemented in Tensorflow and optimized by Adagrad \cite{duchi2011adaptive} for offline evaluation. We make the code publicly available\footnote{https://github.com/rener1199/deep\_memory}.
Online implementation is based on a private system with a distributed MPI cluster.

\textbf{Evaluation Metrics.}
1) \textbf{AUC}: the Area Under the ROC Curve over the test set. The larger the better. 2) \textbf{Logloss}: the value of Eq. (\ref{loss_1}) over the test set. The smaller the better.

\subsection{Effectiveness}
\begin{table}[!t]
\setlength{\tabcolsep}{4pt}
\renewcommand{\arraystretch}{1.2}
\caption{Test AUC and Logloss on two large-scale datasets (no dropout is performed).}
\vskip -8pt
\label{tab_auc}
\centering
\begin{tabular}{|l|c|c||c|c|}
\hline
 & \multicolumn{2}{|c||}{\textbf{Avito (full)}} & \multicolumn{2}{|c|}{\textbf{Company}} \\
\hline
\textbf{Model} & AUC & Logloss & AUC & Logloss\\
\hline
LR & 0.7374 & 0.04328 & 0.6251 & 0.2582 \\
FM & 0.7844 & 0.03990 & 0.6458 & 0.2490 \\
DNN & 0.7939 & 0.03954 & 0.6601 & 0.2379 \\
W\&D & 0.7951 & 0.03948 & 0.6629 & 0.2355 \\
\hline
MA-DNN & \textbf{0.7968} & \textbf{0.03933} & \textbf{0.6703} & \textbf{0.2331} \\
MA-W\&D & \textbf{0.7976} & \textbf{0.03929} & \textbf{0.6710} & \textbf{0.2323} \\
\hline
\end{tabular}
\vskip -8pt
\end{table}

Table \ref{tab_auc} lists the AUC and Logloss values of different models. It is observed that on both datasets, memory augmented models outperform all the baselines. Moreover, MA-DNN outperforms DNN and MA-W\&D outperforms W\&D. These results show that the memory component can indeed extract useful information from users' historical behaviors and improve the prediction performance.

\subsection{Online A/B Test}
We conducted online experiments in an A/B test framework over 6 days in April 2019. The benchmark model is W\&D. Online CTR is defined as the number of clicks over the number of ad impressions. A larger online CTR indicates the enhanced effectiveness of a CTR prediction model. 
Online A/B test results in Table \ref{tab_online} show that MA-W\&D outperforms W\&D and results in an average increase of daily CTR of 2.56\% and an average increase of daily CPM of 0.74\%.

\section{Related Work}

CTR prediction has attracted lots of attention from both academia and industry \cite{he2014practical,cheng2016wide,he2017neural}.
Generalized linear models, such as Logistic Regression (LR) \cite{richardson2007predicting} and Follow-The-Regularized-Leader (FTRL) \cite{mcmahan2013ad}, have shown decent performance in practice. However, a linear model lacks the ability to learn sophisticated feature interactions. Factorization Machines (FMs) \cite{rendle2010factorization} are proposed to model pairwise feature interactions and they show improved performance. 

In recent years, Deep Neural Networks (DNNs) are exploited for CTR prediction and item recommendation in order to automatically learn feature representations and high-order feature interactions \cite{covington2016deep,zhang2016deep,wang2017deep,he2017neural}.
Qu et al. \cite{qu2016product} propose the Product-based Neural Network where a product layer is introduced between the embedding layer and the fully connected layer. Cheng et al. \cite{cheng2016wide} propose Wide\&Deep, which combines LR and DNN to improve both the memorization and generalization abilities of the model. 
Guo et al. \cite{guo2017deepfm} propose DeepFM, which models low-order feature interactions like FM and models high-order feature interactions like DNN. 
To capture dependency on users' sequential behaviors, Zhang et al. \cite{zhang2014sequential} propose Recurrent Neural Network (RNN) based models for CTR prediction. Nevertheless, the application of RNNs to practical CTR prediction services is rather complex.

In this paper, we propose Memory Augmented Deep Neural Network (MA-DNN) for CTR prediction. The proposed MA-DNN achieves a good compromise between DNN and RNN. We are aware of recent work like \cite{chen2018sequential} that also utilizes memory networks \cite{graves2014neural}. However, \cite{chen2018sequential} is proposed for recommender systems and our way of designing the user memory component is different from that in \cite{chen2018sequential}.

\begin{table}[!t]
\setlength{\tabcolsep}{4pt}
\renewcommand{\arraystretch}{1.2}
\caption{Online A/B test results.}
\vskip -8pt
\label{tab_online}
\centering
\begin{tabular}{|l|r|r|}
\hline
\textbf{Model} & CTR gain & CPM gain\\
\hline
W\&D & 0\% & 0\% \\
MA-W\&D & +2.56\% & +0.74\% \\
\hline
\end{tabular}
\vskip -8pt
\end{table}

\section{Conclusion}
In this paper, we propose Memory Augmented DNN (MA-DNN) for CTR prediction in advertising systems. In particular, we create two external memory vectors for each user, memorizing high-level abstractions of what a user possibly likes and dislikes. The proposed MA-DNN achieves a good compromise between DNN and RNN. Actually, the memory component can be augmented to other models as well (e.g., the Wide\&Deep model). Both offline and online experiments demonstrate the effectiveness of memory augmented neural networks for practical CTR prediction services.

\bibliographystyle{ACM-Reference-Format}
\bibliography{ref}

\end{document}